\documentclass[12pt]{article}
\usepackage{mayahq}
\usepackage[utf8]{inputenc}
\usepackage{graphicx}
\usepackage{pdflscape}
\usepackage{float}
\usepackage{amsmath}
\usepackage{amsthm}
\usepackage{amssymb}
\usepackage{scrtime}
\usepackage[dvipsnames]{xcolor}
\usepackage{hyperref}
\usepackage{fnpct}
\usepackage{titletoc,tocloft}
\usepackage[frozencache=true,cachedir=minted-cache]{minted}
\usepackage{booktabs}
\usepackage[backend=biber,style=authoryear,sorting=ynt,natbib=true]{biblatex}

\newcommand{\clqm}{\href{https://github.com/ahgamut/cliquematch}{\texttt{cliquematch}~}}

\newtheorem{definition}{Definition}
\hypersetup{
	colorlinks=true,
	urlcolor=MidnightBlue,
	citecolor=PineGreen,
	linkcolor=MidnightBlue,
}

\author{Gautham Venkatasubramanian \\ \texttt{ahgamut@gmail.com}}

\title{\texttt{cliquematch}: Finding correspondence via cliques in large graphs}

\begin{document}

\maketitle

\begin{abstract}
	The maximum clique problem finds applications in computer vision, bioinformatics, and
	network analysis, many of which involve the construction of correspondence graphs to find
	similarities between two given objects. \clqm is a \texttt{Python} package designed
	for this purpose: it provides a simple framework to construct correspondence graphs, and
	implements an algorithm to find and enumerate maximum cliques in \texttt{C++},
	that can process graphs of a few million edges on consumer hardware, with comparable
	performance to publicly available methods.
\end{abstract}

\tableofcontents

\newpage

\section{Introduction}%
\label{sec:introduction}
Given an undirected graph $G$, a subgraph $H$ of
$G$ is a clique if an edge exists between any two vertices in
$H$. A clique in $G$ is a \textit{maximum clique}
if there exist no cliques of a larger size. The maximum clique problem
\citep{bomze1999} is a special case of the \textit{maximal clique} problem: a clique
is maximal if it is not properly contained in any other clique, therefore all maximum
cliques are also maximal. It is also related to the problem of clique enumeration.

Finding cliques in a graph is applicable to a variety of domains, such as bioinformatics,
robotics, forensics, image analysis  etc. \citep{conte2004}. The applications
transform into a clique problem in two general ways. In \citep{pradalier2003}, vertices
of the graph refer to elements of a dataset, and an \textit{edge indication function} computes a
relationship between every pair of elements. Alternatively, in \citep{horaud1989}, a
\textit{correspondence graph} is constructed to find similar substructures between two different
objects; vertices correspond to potential mappings between similar elements of the two
objects, and edges reinforce the mappings.

\clqm \footnote{\url{https://github.com/ahgamut/cliquematch}} is a \texttt{Python} package designed to construct
correspondence graphs and find maximum cliques. It implements a modified version of a
well-known maximum clique algorithm \citep{pattabiraman2015} in \texttt{C++}, and
uses template programming to provide a simple framework for constructing correspondence
graphs. \clqm makes use of the \texttt{pybind11} library to provide
\texttt{Python} bindings. The remainder of this section provides a simple example
using \clqm. \autoref{sec:mcs} describes the algorithm used by \clqm when finding
maximum cliques, \autoref{sec:correspondence} shows how problems of data association can be
converted into finding maximum cliques in a graph. \autoref{sec:applications} provides
examples showcasing how \clqm can be used to solve such kinds of problems.
\autoref{sec:conclusions} summarizes the properties of \clqm and discusses future
directions.


\subsection{Basic Usage}%
\label{sub:basic}

The core functionality of \clqm involves loading an undirected graph and finding a
clique. The graph can be loaded from edge lists, adjacency matrices, adjacency lists, and
text files that follow the Matrix Market Coordinate Text File format.
\footnote{\url{https://math.nist.gov/MatrixMarket/formats.html}}

\begin{minted}[autogobble,mathescape,linenos,numbersep=5pt]{python}
	import cliquematch 
	G = cliquematch.Graph.from_file("cond-mat-2003.mtx")
	print(G) # cliquematch.core.Graph object at 0x559e7da730c0
	# (n_vertices=31163,n_edges=120029,lower_bound=1,upper_bound=4294967295,
	# time_limit=1,use_heuristic=False,use_dfs=True,search_done=False) 
	G.get_max_clique() 
	# [9986, 9987, 10066, 10068, 10071, 10072, 10074, 10076, 
	# 10077, 10078, 10079, 10080, 10081, 10082, 10083, 10085, 
	# 10287, 10902, 10903, 10904, 10905, 10906, 10907, 10908, 10909]
\end{minted}

The search for a clique can be modified by: (a) setting the bounds on clique size (via
\texttt{lower\_bound} and \texttt{upper\_bound}), (b) choosing to use the heuristic
method, the depth-first search, or both (via \texttt{use\_heuristic} and
\texttt{use\_dfs}), and (3) setting a time limit for the search (via
\texttt{time\_limit}).

The search for maximum cliques can be resumed and interrupted intermittently using
\texttt{search\_done} and \texttt{get\_max\_clique()} in a loop, which is useful for
incremental searching in the case of dense graphs. \texttt{reset\_search()} resets the
search for maximum cliques in case different bounds are required.

\begin{minted}[autogobble,mathescape,linenos,numbersep=5pt]{python}
G.reset_search() 
while not G.search_done: 
    answer = G.get_max_clique(
        lower_bound=1, upper_bound=1729,
        use_heuristic=True, use_dfs=True,
        time_limit=100, continue_search=True
    )
\end{minted}

The \texttt{all\_cliques()} method can be used to obtain all cliques of a particular size
from the graph $G$. \texttt{all\_cliques()} does not find all the
cliques at once; the cliques are discovered upon the user's repeated requests.

\begin{minted}[autogobble,mathescape,linenos,numbersep=5pt]{python}
import cliquematch
G = cliquematch.Graph.from_file("cond-mat-2003.mtx")
for clique in G.all_cliques(size=24):
     print(clique)
\end{minted}

\section{The Maximum Clique Problem - A Literature Review}%
\label{sec:mcs}

The maximum clique problem is NP-Hard \citep{garey1979computers}, and many algorithms for
computing an exact solution have been discovered. These usually involve a possible
optimal vertex ordering, fast heuristic bounds on the maximum clique size, followed by
\textit{branch-and-bound}: performing a depth-first search from each vertex to find
cliques, and pruning the search space to avoid unnecessary calculations. The earliest
such algorithm \citep{carraghan1990} sorts vertices in ascending order of degree, with
search steps being pruned if they cannot beat the current maximum.
\cite{ostergaard2002} sorts vertices in descending order, and processes them in a
defined sequence for better performance. MCQ \citep{tomita2003} first sorts vertices
in descending order, and uses an approximate coloring for additional sorting of the
vertices, which also helps pruning in the clique.  A later version
\citep{tomita2010} improves on the approximate coloring used so as to maximize
pruning.

More recent algorithms for finding maximum cliques focus on massive sparse graphs; these
may require specialized hardware, and attempt to use the parallel nature of the problem.
FMC \citep{pattabiraman2015} prunes vertices with degree less than the current maximum
clique size as early as possible, and ignores vertices that have already been processed;
it also provides a degree-based heuristic method to obtain a lower bound on the maximum
clique size. PMC \citep{rossi2015} uses the core-number \citep{seidman1983network} of
a vertex instead of the degree, which provides a tight lower and upper bound, thereby
pruning the search space more effectively, and provides a parallel-friendly
implementation based on \texttt{OpenMP} \citep{openmp1998}.  BBMC
\citep{segundo2011} uses bitstrings of 64-bit machine words to encode the adjacency
matrix and vertex sets, to benefit from bit-parallelism in set operations. BBMCSP
\citep{segundo2016} defines a sparse encoding for bitstrings; it also unrolls the
initial search step to avoid unnecessary recursive calls. It is interesting to note that
pruning methods based on heuristics are not optimal for some kinds of real-world graphs;
RMC \citep{lu2017} describes a probabilistic algorithm for finding maximum
cliques, with examples and benchmarks showcasing potential limitations in heuristic-based
methods.

The algorithm used in \clqm is mostly similar to FMC: the depth-first search and the
heuristic method both filter out vertices based on degree, and the search space is pruned
based on potential to beat the current maximum. \clqm also uses bitstrings compressed
into 32-bit machine words, similar to BBMC, to represent the vertex sets during the
clique search. However, \clqm differs from FMC in three ways:

\begin{itemize}
	\item Instead of filtering out completed vertices (FMC Pruning 2) and neighbors of a vertex
	      $v$ with lesser degree than the current maximum clique size (FMC
	      Pruning 3), \clqm filters out all neighbors of $v$ of lesser degree
	      than $d(v)$, the degree of $v$.  This means that every
	      maximum clique is now found using the vertex of least degree and the search is now
	      amortized over all the vertices, which reduces reliance on vertex ordering.
	\item The heuristic method returns a clique instead of just the lower bound. This helps users
	      to obtain a clique quickly in case the branch-and-bound method is too slow.
	\item The branch-and-bound method is repurposed to also provide clique enumeration. This allows
	      to find all cliques of a given size, as there might be multiple maximum cliques. The
	      clique enumeration is done in a lazy manner; new cliques are found incrementally upon
	      request.
\end{itemize}

The performance of \clqm on various benchmark graphs is comparable to existing
\texttt{C++} implementations, see  \autoref{sec:performance}.

\section{Correspondence Graphs}%
\label{sec:correspondence}
Finding maximum cliques in graphs can be applied to data association problems, where the
aim is to find similarity between two objects by comparing their components. Such
problems are found in bioinformatics \citep{gardiner1997}, robotics
\citep{pradalier2003}, forensics \citep{fingerprint1999}, image analysis
\citep{horaud1989} etc. These problems can be solved by constructing a
\emph{correspondence graph}, a general form of the association graph \citep{kozen1978clique}
used for subgraph isomorphism. Given two given objects $P$ and
$Q$, the correspondence graph $G_{P,Q}$ constructs the
largest possible correspondence between $P$ and
$Q$ by extending mappings in a pairwise manner.

\begin{figure}[htpb]
	\centering
	\hspace*{-1in}
	\includegraphics[width=1.4\linewidth]{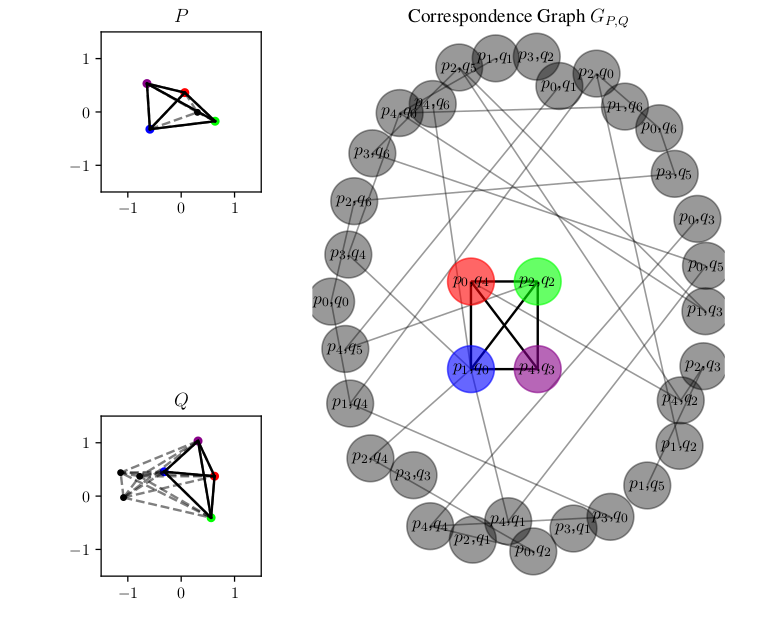}
	\caption{A sample correspondence graph for sets of points in two dimensions. The corresponding
		points are marked in red, green, blue, and purple. The vertices of the correspondence graph $G$ which refer to
		these are marked with the same colors. Note that the edges between the colored vertices
		form a maximum clique, and the configurations of the corresponding points (thicker lines) are the same in both
		$P$ and $Q$.}
	\label{fig:corrgraph}
\end{figure}

\begin{definition}[Correspondence Graph]
	Let $P = \{p_1, p_2, .... p_M\} $ and $Q = \{q_1, q_2, .... q_N\}$ be two
	sets of elements of length $M$ and $N$
	respectively. Let $G (V,E)$ be an undirected graph, where
	$V = P \times Q$.  An edge $e \in E$ is drawn between
	$v_1(p_{i\_1}, q_{j\_1})$ and $v_2(p_{i\_2}, q_{j\_2})$ $\in V$ iff for a given
	boolean function $f : (P \times P \times Q \times Q) \rightarrow \{0,1\} $
	\begin{align}
		\label{eq:eq1}
		(v_1, v_2) \in E \iff
		f(p_{i\_1}, p_{i\_2}, q_{j\_1}, q_{j\_2}) = 1;\ ~~ &
		i_1 \ne i_2,\  j_1 \ne j_2
	\end{align}
\end{definition}

Finding a maximum clique $H$ in $G$ is equivalent
to finding the largest correspondence between $P$ and
$Q$ as shown in the following step by step argument.
\begin{description}
	\item (a) $V_H \subset V_G = P \times Q$.
	\item (b) Let $P' \subset P$, and $Q' \subset Q$ be such that
	      $$
		      (p_i,q_j) \in V_H \iff p_i \in P', q_j \in Q'.
	      $$
	\item (c) $H$ is a clique, so there exists an edge between every pair of
	      vertices in $V_H$. Remember that an edge can be drawn only if
	      Equation~\ref{eq:eq1} is satisfied, therefore
	      \begin{align*}
		                                 & f(p_{i\_1}, p_{i\_2}, q_{j\_1}, q_{j\_2}) = 1, &
		      i_1 \ne i_2,\  j_1 \ne j_2 & ~~
		      \forall\  (~v_1(p_{i\_1}, q_{j\_1}),~v_2(p_{i\_2}, q_{j\_2})~) \in V(H)       \\
		      \implies                   & f(p_{i\_1}, p_{i\_2}, q_{j\_1}, q_{j\_2}) = 1, &
		      i_1 \ne i_2,\  j_1 \ne j_2 & ~~
		      \forall\  p_{i\_1}, p_{i\_2} \in P';
		      \forall\  q_{j\_1}, q_{j\_2} \in Q'
	      \end{align*}
	      because every vertex in $V_H$ is a pair of elements, one from
	      $P'$ and the other from $Q'$. Hence, there exists a
	      correspondence between $P'$and $Q'$.
	\item (d) $H$ is a maximum clique, so there exists no clique in
	      $G$ that is larger than $H$. Thus,
	      $P'$ and $Q'$ are subsets of $P$
	      and $Q$ having the largest possible correspondence.
\end{description}

Note that $f$ requires two pairs of elements $p_{i\_1}, p_{i\_2},
	q_{j\_1}, q_{j_2}$,
when constructing an edge of the correspondence graph, and thus there is a pairwise
correspondence between elements of $P'$ and $Q'$.
$f$ can be optimized to benefit from properties of
$P$ and $Q$. A common use case is if
$P$ and $Q$ are point-clouds in an
$n$-dimensional space (see \autoref{fig:corrgraph}), the function
$f$ can be:

$$
	f(p_{i\_1}, p_{i\_2},
	q_{j\_1}, q_{j\_2}) = 1 \iff ||
	d_P(p_{i\_1}, p_{i\_2}) -
	d_Q(q_{j\_1}, q_{j\_2}) || \leq \epsilon
$$

where $d_P$ is a distance metric on $P$,
$d_Q$ is a distance metric on $Q$, and
$\epsilon$ is a small positive real number. Therefore, the edge construction
rule in Equation \ref{eq:eq1} is modified to:

\begin{align}
	\label{eq:eq2}
	(v_1, v_2) \in E \iff || d_P(p_{i\_1}, p_{i\_2}) -
	d_Q(q_{j\_1}, q_{j\_2}) || \leq \epsilon ~~ &
	i_1 \ne i_2,\  j_1 \ne j_2 ~~
\end{align}

\section{Applications}%
\label{sec:applications}

\clqm can construct correspondence graphs where $P, Q$ are either 2D
\texttt{numpy} arrays or \texttt{Python} lists, via the below classes:
\begin{itemize}
	\item \texttt{A2AGraph}, where $P$ and $Q$ are
	      \texttt{numpy} arrays
	\item \texttt{L2LGraph}, where $P$ and $Q$ are lists
	      of arbitrary objects, and
	\item \texttt{A2LGraph} and \texttt{L2AGraph}, for cases that may require mapping a
	      list of objects to \texttt{numpy} arrays of related data.
\end{itemize}

The user is required to define the function $f$ or the metrics
$(d_P, d_Q)$ for \clqm to perform the construction of the graph.  \clqm uses
\texttt{pybind11} for \texttt{Python} wrappers, so one can define
$d_P$, $d_Q$, and $f$ as regular
\texttt{Python} functions or \texttt{Callable} objects for fast
prototyping. Note that accessing elements of $P$ and
$Q$ is done only within these functions.

\begin{minted}[autogobble,mathescape,linenos,numbersep=5pt]{python}
def euclidean(P, i1, i2): 
    return sqrt(sum((P[i1]- P[i2]) ** 2))

class MyCustomCondition(object):
    def __call__(P, i1, i2, Q, j1, j2): 
        if my_condition_works: 
            return True 
        return False
\end{minted}

Once $G$ has been constructed as per the given conditions, \clqm
searches for cliques -- the search parameters can be defined as per
\autoref{sub:basic} -- and returns the subsets with largest correspondence, as seen
in the following examples.

\subsection{Image Registration and Matching using interest points}%
\label{sub:imregister}
Image registration can be converted into point-cloud registration by selecting a suitable
function to obtain interest points, following which simple distance metrics can be used
to construct a correspondence graph as in Equation~\ref{eq:eq2}. Once a
maximum clique has been found, the sets of corresponding points can be used obtain a
matching score, or perform a registration of the image.

CCMM \citep{segundo2015} performs feature matching for color images by computing SURF
descriptors \citep{bay2006} to obtain interest points. The algorithm to construct
a correspondence graph $G$ can be described as follows:

\begin{itemize}
	\item $P$, $Q$ are the sets of SURF keypoints in the
	      first and second images.
	\item $d_P$ and $d_Q$ are the Euclidean metrics.
	\item Additionally, apply a condition function $f$ that allows an edge from
	      $(p_{i\_1}, q_{j\_1})$ to $(p_{i\_2}, q_{j\_2})$ if and only if $q_{j_1}$ is
	      one of the top $k$ SURF descriptor matches of $p_{i_1}$
	      and $q_{j_2}$ is one of the top $k$ SURF descriptor
	      matches of $p_{i_2}$; where $k$ is some integer.
	\item The correspondence graph $G$ is constructed using the distance metrics
	      and $f$, for some appropriate values of $k$ and
	      $\epsilon$.
\end{itemize}

\begin{figure}[htbp]
	\centering
	\includegraphics[width=0.9\linewidth]{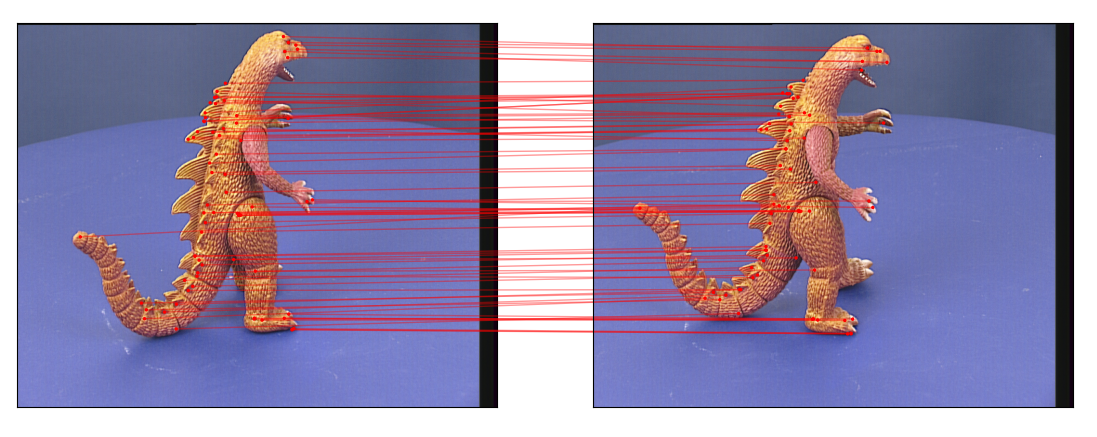}
	\caption{The CCMM algorithm tested on images from the Dinosaur dataset \citep{vggdatasets}. The
		points in red are the corresponding points. }
	\label{fig:ccmm}
\end{figure}

An implementation of the above steps using the \texttt{cliquematch.A2AGraph} and
\texttt{OpenCV} is available on Github \footnote{\url{https://github.com/ahgamut/cliquematch/blob/master/examples/ccmm.py}}, a sample result
is shown in Figure \ref{fig:ccmm}. A similar procedure can be followed for
registering or matching a pair of images, based on the kinds of interest points
extracted:

\begin{itemize}
	\item \citep{fingerprint1999} compares fingerprint images by selecting corresponding pairs of
	      minutia: the vertices of the correspondence graph are mappings of minutia, and edges are
	      drawn with respect to a function computing angle, distance, and ridge counts.
	\item \citep{park2020} extracts SURF points and computes maximum cliques on a
	      correspondence graph to perform alignment of footwear outsole impressions.
	\item \citep{theiler2012} performs registration of laser scans by computing tie points, and
	      uses description vectors for each point along with the Euclidean distance metric to
	      ensure construction of a sparse graph. The correspondence computed via finding maximum
	      cliques is used to register the scans.
\end{itemize}

\subsection{Matching of Molecular Structures}%
\label{sub:protein}

The structure of molecules can be represented as an attributed graph, and therefore
matching the 3-D structures of two different molecules can be converted into finding a
clique in their correspondence graph. \citep{gardiner1997} provides a procedure for
structure matching of molecules via correspondence graph, which can be described as
follows\footnote{Implementation available at \url{https://github.com/ahgamut/cliquematch/blob/master/examples/molecule.py}}:

\begin{itemize}
	\item $P$, $Q$ are the sets of atoms in the first and
	      second molecules to be matched.
	\item $d_P$ and $d_Q$ are the Euclidean metrics to measure
	      inter-atomic distances.
	\item Additionally, apply a condition function $f$ that allows an edge if
	      and only if a bond exists between the pairs of atoms being mapped.
	\item The correspondence graph $G$ is constructed using the distance metrics
	      and $f$, for some appropriate value of $\epsilon$.
	      $f$ can be modified to account for additional properties (e.g. match
	      ring bonds to ring bonds, valence).
\end{itemize}

\begin{figure}[phtb]
	\centering
	\includegraphics[width=0.9\linewidth]{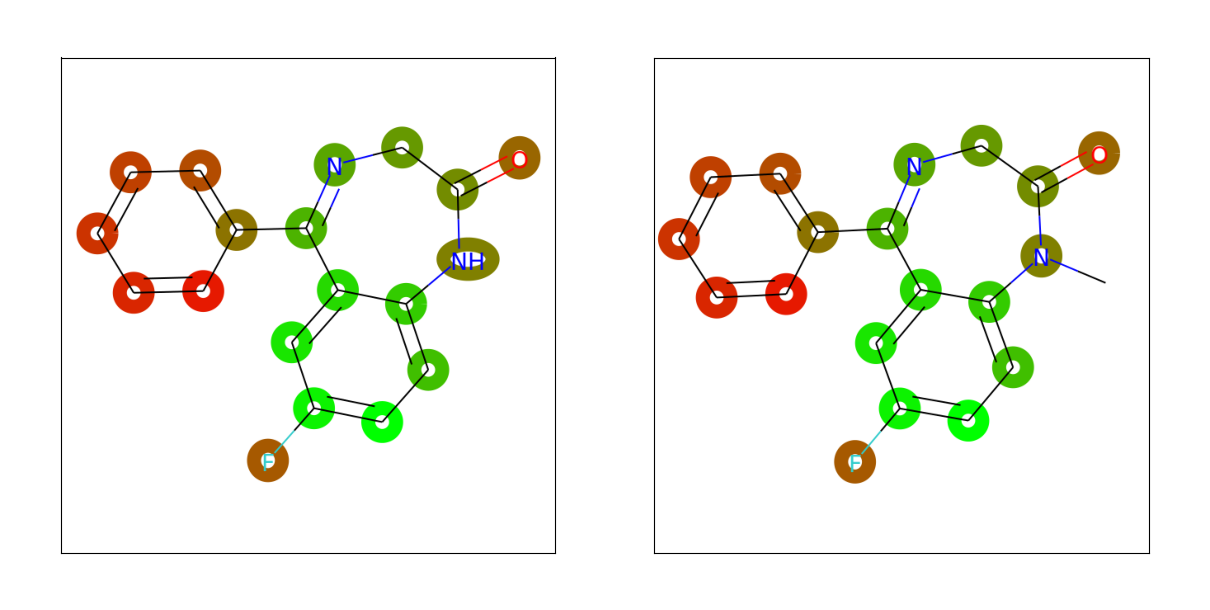}
	\caption{Molecule structure matching using inter-atomic distances. The molecules were obtained
		from the datasets provided in \citep{sutherland2003}.}
	\label{fig:molecule}
\end{figure}

An illustrative example using \clqm is shown in Figure \ref{fig:molecule}. A similar
procedure is followed for matching protein molecules using their secondary structure
elements (SSEs). \citep{butenko2006} gives an overview of applying clique-based
methods in biochemistry.

\subsection{Subgraph Isomorphism}%
\label{sub:isograph}

Finding a subgraph isomorphism between two graphs can be solved by constructing a
correspondence graph as described in \citep{kozen1978clique}:

\begin{itemize}
	\item Let $G_P(P, E_P)$ and $G_Q(Q, E_Q)$ be simple, unweighted, undirected
	      graphs such that $G_Q$ is isomorphic to a subgraph of
	      $G_P$. The vertices of the graphs are sets $P$ and
	      $Q$.
	\item Define a correspondence graph $G^*(V,E)$ where $V = P \times Q$.
	\item Define a boolean condition function $f$ to construct edges in
	      $G^*$ as below: $$ f(P, i_1, i_2, Q, j_1, j_2) = 1 \iff (q_{j_1}, q_{j_2}) \in E_Q \implies (p_{i_1}, p_{i_2}) \in
		      E_P $$
	\item Once $G^*$ has been constructed, finding a maximum clique will give the
	      vertices of the isomorphic subgraphs.
\end{itemize}

\clqm provides an \texttt{IsoGraph} class which encapsulates the above
functionality.

\section{Conclusions and Future Work}%
\label{sec:conclusions}
I have described the capabilities of \clqm, a \texttt{Python} package that finds
maximum cliques in large sparse graphs and shown that its performance is comparable to
other publicly available methods. I also provided examples showing that the
implementation of \clqm can be used to in solving data association problems in different
domains, by constructing a correspondence graph and finding a maximum clique.

Multiple aspects of \clqm can be developed further: the core search algorithm can be
modified to find maximum cliques in a weighted graph. The computation time can improved
with better heuristics, like vertex coreness and approximate coloring. The construction
of correspondence graphs is applicable to many problem domains. There is also scope for
clique augmentation or lenient matching methods (para-cliques) using the \clqm design,
and for providing GPU or streaming-specific implementations.

\section*{Acknowledgments}
Some of this work was done at the National Institute of Standards and Technology (NIST)
in support of the Forensic Footwear Research Project. I would like to thank Dr. Martin
Herman, Dr. Steve Lund, and Dr. Hari Iyer of NIST for helpful discussions.

\clearpage
\nocite{python}
\nocite{numpy}
\nocite{pybind11}
\nocite{eigen2014}
\nocite{openmp1998}
\nocite{opencv}

\printbibliography

\newpage

\begin{appendix}
	\savegeometry{default}
	\newgeometry{,hmargin=1.0cm,vmargin=0.5cm}
	\pagestyle{empty}
	\renewcommand{\arraystretch}{2.0}

	\section{Performance}%
	\label{sec:performance}
	\begin{table}[htb]
		\centering
		\begin{tabular}{|c|c|c|c|c|c|c|c|c|c|c|}
			\toprule
			                 & $|V|$   & $|E|$    & $\omega$ & $t_{cm}$         & $t_{fmc}$       & $t_{pmc}$       & $t_{cm-heur}$ & $\omega_{cm-heur}$ & $t_{fmc-heur}$ & $\omega_{fmc-heur}$ \\
			\midrule
			Erdos02          & 6927    & 8472     & 7        & \textbf{0.0003}  & 0.0013          & 0.0022          & 0.0002        & 6                  & 0.0009         & 7                   \\
			Erdos972         & 5488    & 7085     & 7        & \textbf{0.0002}  & 0.0007          & 0.0015          & 0.0002        & 7                  & 0.0002         & 6                   \\
			Erdos982         & 5822    & 7375     & 7        & \textbf{0.0004}  & 0.0005          & 0.0015          & 0.0002        & 7                  & 0.0002         & 7                   \\
			Erdos992         & 6100    & 7515     & 8        & \textbf{0.0002}  & 0.0004          & 0.0017          & 0.0002        & 8                  & 0.0002         & 8                   \\
			Fault\_639       & 638802  & 14626683 & 18       & 21.3265          & \textbf{14.456} & -               & 1.2731        & 18                 & 2.4945         & 18                  \\
			brock200\_2      & 200     & 9876     & 12       & 0.4408           & 0.6408          & \textbf{0.0018} & 0.0031        & 10                 & 0.0023         & 9                   \\
			c-fat200-5       & 200     & 8473     & 58       & 0.1485           & 0.4204          & \textbf{0.0003} & 0.0004        & 58                 & 0.0106         & 58                  \\
			ca-AstroPh       & 18772   & 198110   & 57       & \textbf{0.0024}  & 0.0802          & 0.0137          & 0.0068        & 57                 & 0.0286         & 57                  \\
			ca-CondMat       & 23133   & 93497    & 26       & \textbf{0.001}   & 0.0048          & 0.0083          & 0.0016        & 26                 & 0.004          & 26                  \\
			ca-GrQc          & 5242    & 14496    & 44       & \textbf{0.0002}  & 0.0005          & 0.0018          & 0.0002        & 44                 & 0.0011         & 44                  \\
			ca-HepPh         & 12008   & 118521   & 239      & \textbf{0.0041}  & 0.0138          & 0.016           & 0.0045        & 239                & 0.2589         & 239                 \\
			ca-HepTh         & 9877    & 25998    & 32       & \textbf{0.0002}  & 0.0007          & 0.0036          & 0.0001        & 32                 & 0.0001         & 32                  \\
			caidaRouterLevel & 192244  & 609066   & 17       & \textbf{0.0672}  & 0.2193          & 0.0784          & 0.0258        & 17                 & 0.0723         & 15                  \\
			coPapersCiteseer & 434102  & 16036720 & 845      & \textbf{0.028}   & 0.8812          & 1.8326          & 0.0501        & 845                & 16.2965        & 845                 \\
			com-Youtube      & 1134890 & 2987624  & 17       & \textbf{2.467}   & 10.6184         & -               & 0.2301        & 16                 & 0.3597         & 13                  \\
			cond-mat-2003    & 31163   & 120029   & 25       & \textbf{0.0024}  & 0.013           & 0.0104          & 0.0032        & 25                 & 0.0054         & 25                  \\
			cti              & 16840   & 48232    & 3        & 0.0283           & 0.0052          & \textbf{0.0048} & 0.0058        & 3                  & 0.0014         & 3                   \\
			hamming6-4       & 64      & 704      & 4        & 0.0018           & \textbf{0.0005} & 6.1989          & 0.0001        & 4                  & 6.6042         & 4                   \\
			johnson8-4-4     & 70      & 1855     & 14       & 0.6535           & 0.1582          & \textbf{0.0003} & 0.0005        & 14                 & 0.0005         & 14                  \\
			keller4          & 171     & 9435     & 11       & \textbf{10.8767} & 15.7847         & -               & 0.0022        & 9                  & 0.0027         & 9                   \\
			loc-Brightkite   & 58228   & 214078   & 37       & 7.0798           & 2.9293          & \textbf{0.0271} & 0.0077        & 36                 & 0.0155         & 31                  \\
			\bottomrule
		\end{tabular}
		\normalsize
		\caption{Comparing \clqm performance on some benchmark graphs. $|V|$ and
			$|E|$ denote the number of nodes and edges in the graph.
			$\omega$ denotes the size of the maximum clique found. All the
			branch-and-bound methods agreed on the maximum clique size in every benchmark.
			$t_{cm}$, $t_{fmc}$, and $t_{pmc}$ denote the time taken by \clqm, FMC, and PMC respectively in the branch-and-bound search: the least time is in bold text.
			$w_{cm-heur}$, $t_{cm-heur}$ denote the size of clique and time taken by the heuristic method
			in \clqm; and similarly for $w_{fmc-heur}$ and $t_{fmc-heur}$. A minus
			sign (\texttt{-}) indicates that the program returned an error without
			completing the calculation.}
	\end{table}

	The benchmark graphs were obtained from the Stanford SNAP collection
	\citep{snapnets}, the University of Florida Sparse Matrix collection
	\citep{ufsparse}, and the DIMACS Challenges (\cite{dimacs2},
	\cite{dimacs10}). I thank the authors of FMC \footnote{\url{http://cucis.ece.northwestern.edu/projects/MAXCLIQUE/download.html}} and PMC
	\footnote{\url{https://github.com/ryanrossi/pmc}} for making their source code publicly available.

	I used \texttt{gcc 7.5.0} to compile the programs at optimization level
	\texttt{-O3}. For \clqm I set the \texttt{BENCHMARKING} flag to
	\texttt{1} before compilation. I compiled and tested the programs on a 64-bit
	\texttt{Ubuntu 18.04} system with \texttt{Intel(R) Core(TM) i5-4200U CPU @ 1.60GHz} and \texttt{4GB RAM}. The
	following command line parameters were used:
	\begin{itemize}
		\item \texttt{fmc -t 0 -p} was used to run the FMC branch-and-bound algorithm.
		\item \texttt{pmc -t 1 -r 1 -a 0 -h 0 -d} (single CPU thread, reduce wait time of 1 second, full algorithm, skip heuristic, search in descending order) was used to run the PMC branch-and-bound algorithm.
		\item \texttt{fmc -t 1} was used to run the FMC heuristic algorithm.
		\item A small python script similar to the code block in \autoref{sub:basic} was used to run
		      the \clqm algorithms.
	\end{itemize}

	\loadgeometry{default}
\end{appendix}
\end{document}